\documentclass[preprint]{aastex631}

\hypersetup{linkcolor=blue,citecolor=blue,filecolor=cyan,urlcolor=blue}

\shorttitle{Foreshock ULF waves at Mars}
\shortauthors{Andrés et al. (2024)}
\graphicspath{{./}{figures/}}

\usepackage{amsmath, ulem}
\usepackage{lipsum}  

\usepackage{soul}

\begin{document}

\title{Foreshock Ultra-Low Frequency Waves at Mars: Consequence on the Particle Acceleration Mechanisms at the Martian Bow Shock}

\correspondingauthor{Nahuel Andr\'es}
\email{nandres@df.uba.ar}
\author[0000-0002-1272-2778]{Nahuel Andr\'es}
\affiliation{CONICET - Universidad de Buenos Aires, Instituto de Física Interdisciplinaria y Aplicada (INFINA), Ciudad Universitaria, 1428 Buenos Aires, Argentina}
\affiliation{Universidad de Buenos Aires, Facultad de Ciencias Exactas y Naturales, Departamento de Física, Ciudad Universitaria, 1428 Buenos Aires, Argentina}
\author[0000-0001-9210-0284]{Norberto Romanelli}
\affiliation{Department of Astronomy, University of Maryland, College Park, MD, USA}
\affiliation{Planetary Magnetospheres Laboratory, NASA Goddard Space Flight Center, Greenbelt, MD, USA}
\author[0000-0001-5332-9561]{Christian Mazelle}
\affiliation{Institut de Recherche en Astrophysique et Planétologie, CNRS, Université Paul Sabatier, CNES, Toulouse, France}
\author[0000-0002-4768-189X]{Li-Jen Chen}
\affiliation{Geospace Physics Laboratory, NASA Goddard Space Flight Center, Greenbelt, MD, USA}
\author[0000-0002-1215-992X]{Jacob R. Gruesbeck}
\affiliation{Planetary Magnetospheres Laboratory, NASA Goddard Space Flight Center, Greenbelt, MD, USA}
\author[0000-0002-6371-9683]{Jared R. Espley}
\affiliation{Planetary Magnetospheres Laboratory, NASA Goddard Space Flight Center, Greenbelt, MD, USA}

\begin{abstract}

Using Mars Atmosphere and Volatile EvolutioN Magnetometer observations, we report the first statistical study of ultra-low frequency (ULF) waves at the Martian foreshock. The analyzed foreshock ULF wave events are observed in the 0.008-0.086 Hz frequency range, with nearly circular and elliptical left-handed polarization in the spacecraft reference frame. These waves are propagated quasi-parallel to the ambient magnetic field, with a moderate wave amplitude. All these properties are consistent with fast magnetosonic waves, most likely generated through the ion-ion right-hand resonant instability. In addition, our results suggest that the associated resonant backstreaming protons' velocities parallel to the mean magnetic field in the solar wind reference frame is $1.33\pm0.40$ times the solar wind velocity. The similarity between our results and previous reports at other foreshocks may indicate the presence of a common acceleration process acting in planetary bow shocks and that is responsible for this particular backstreaming population. 

\end{abstract}

\section{Introduction}\label{sec:intro}

The foreshock is the magnetically connected spatial region located upstream of a planetary bow shock. Within this region, there are at least two distinct particle populations, the solar wind particles coming from the Sun, and a smaller group of backstreaming particles, which are generated either by the reflection of particles from the planet's bow shock or by the leakage particles from the planet's magnetosheath \citep[e.g.,][]{E2005}. The interaction between these populations can lead to various plasma instabilities \citep[see,][]{G1993}, resulting in the excitation of diverse plasma wave modes \citep{B1997}. Notably, different backstreaming ion distributions, i.e., field-aligned, intermediate or diffuse distributions, can generate different plasma waves types. These include quasi-monochromatic “30 s” waves (which have a period of approximately 30 seconds in the spacecraft reference frame at the Earth's foreshock), shocklets (i.e., waves with steepened edges), the “1 Hz” whistler waves, or large amplitude waves (sometimes referred to as SLAMS, short for short, large amplitude magnetic structures)  \citep[see,][]{F1974,Go1978,P1979,P1981,H1983,E2005}. In situ observations from different planetary foreshocks indicate that quasi-monochromatic Ultra-Low Frequency (ULF) waves are left-handed in the spacecraft frame, propagating towards the Sun (in the solar wind rest frame) and away from the shock \citep{H1981,E2005}. Consequently, these waves have been identified as right-handed in the plasma frame and are consistent with excitation by backstreaming protons through the ion-ion right hand resonant beam instability \citep[see,][and reference therein]{G1993}.

Understanding the characteristics of ULF waves in planetary foreshocks is crucial for identifying and comparing physical processes, as well as {\color{black} characterizing the} particle acceleration mechanisms occurring at planetary bow shocks \citep[e.g.,][]{H1982,E2005,S2016,S2018,M2000,M2003,A2013,A2015,R2020,R2021}. Specifically, \citet{H1982} investigated the occurrence of foreshock waves at Venus, Earth and Jupiter, using various spacecraft in situ measurements. The authors found an empirical relationship between the magnitude of the interplanetary magnetic field and the observed frequency of these waves. Using Venus Express observations, \citet{S2016} conducted a statistical study to identify and characterize ULF waves at the Venus foreshock. Their findings indicated properties of the quasi-monochromatic ULF waves similar to those found in the terrestrial foreshock \citep[see][]{E2005}. Additionally, \citet{S2018} identified the ULF foreshock wave boundary, suggesting that the backstreaming ions upstream of the Venus bow shock are the primary energy source of these ULF waves. \citet{R2020} performed a statistical analysis of the main properties of ULF waves at the Hermean foreshock using MErcury Surface, Space ENvironment, GEochemistry, and Ranging (MESSENGER) magnetometer observations. They also found polarization properties similar to those of the quasi-monochromatic “30 s” waves observed at Earth's foreshock, but with a smaller normalized wave amplitude. Moreover, they estimated the speed of resonant backstreaming protons in the solar wind reference frame.

In contrast with other planets, the task of identifying and characterizing the quasi-monochromatic ULF waves on Mars's foreshock is significantly more complex due to the concurrent existence of another type of waves known as Proton Cyclotron Waves (PCWs). Due to the lack of an intrinsic global magnetic field, the Martian exosphere is extended beyond the bow shock and is subject to various ionizing mechanisms \citep[see,][]{Ch2014}, generating a population of newborn planetary protons \citep[e.g.,][]{Y2015,R2017}. The interaction between these newborn protons population and the incoming solar wind population is responsible for the generation of waves observed at a frequency near the local proton cyclotron frequency $f_{ci}$ in the spacecraft reference frame \citep[e.g.,][]{R1990,B2002,M2004,R2013,Ro2016,Ru2015,Ru2016,L2020,A2020,Romeo2021}. Since these waves have similar properties in the spacecraft frame, such as frequency, polarization, and propagation angle with respect to the ambient magnetic field, distinguishing between foreshock ULF waves and PCWs requires a specific methodology. Indeed, to the best of our knowledge, there {\color{black} has been} only one reported event of foreshock ULF wave activity at Mars. Using magnetic field and plasma observations from the Mars Atmosphere and Volatile EvolutioN (MAVEN) mission, \citet{Sh2020} reported an event of ULF waves when MAVEN was connected to the Martian bow shock. Spectral analysis revealed that the frequency of this wave event was approximately 0.040 Hz in the spacecraft frame, approximately a factor two lower than the local proton cyclotron frequency $f_{ci} \sim$ 0.088 Hz. Additionally, the authors reported a propagation angle of about 34$^\circ$ relative to the mean magnetic field, with a left-hand elliptical polarization in the spacecraft frame, and a phase speed much lower than the solar wind velocity. Numerically, \citet{J2022} investigated the solar wind interaction with Mars using a global three-dimensional hybrid model. The authors found a well-developed, extensive ion foreshock and the presence of large-scale ULF waves in two specific regions. In the so-called near region, the wave period is 71–83 s and the propagation is nearly perpendicular to the magnetic field with both left- and right-handed wave polarization. On the other hand, a far region where the wave period is 25–28 s, the propagation angle varies between 20$^\circ$ and 50$^\circ$, and the waves present a left-handed polarization with respect to the ambient magnetic field.


The main objective of the present work is to identify the Martian quasi-monochromatic ULF foreshock waves, separate them from the PCW activity and conduct a statistical analysis of the main wave properties. Our investigation is organized as follows: in Section \ref{sec:obs} we briefly describe the MAVEN observations and the criteria used to identify the foreshock ULF wave activity and the wave properties. In Section \ref{sec:case}, we characterized in detail an event identified as a foreshock ULF waves. In Section \ref{sec:stat} we reported our main statistical results, i.e., we reported the frequency, polarization and propagation angles of these type of waves. Moreover, we report particle velocity ratios derived from the acceleration mechanism at the Martian bow shock compatible with the ULF waves properties. Finally, in Section \ref{sec:diss} we present our main discussions and conclusions. 

\section{MAVEN MAG Observations and Detection Criteria} \label{sec:obs}

The MAVEN mission was launched on November 18, 2013, and entered Mars' orbit on September 21, 2014. Its primary objectives are to investigate Mars' upper atmosphere, ionosphere, and their interactions with the solar wind. MAVEN's orbit features a nominal periapsis altitude of 150 km, an apoapsis altitude of 6220 km, and an orbital period of approximately 4.5 h \citep{J2015}. The significant apoapsis altitude enable the investigation of plasma properties upstream of Mars bow shock, particularly in the Martian foreshock region. In the present work, we have analyzed MAVEN MAG observations with 32 Hz sampling cadence \citep{Con2015} from November 17, 2014 until November 30, 2016 upstream from the Martian bow shock. {\color{black} To identify ULF waves in Mars' foreshock region, first we analyzed time intervals when MAVEN was magnetically connected to the Martian bow shock. To achieve this, we separated the time series into 255-s events and calculated the foreshock coordinates \citep{Greenstadt1986} using the spacecraft's position, the local mean magnetic field, and a Martian bow shock model \citep{Gru2018}. This technique has been extensively used in the literature to estimate the spacecraft location with respect to a planetary foreshock \citep[e.g.,][]{M1998,A2013,A2015,K2017,R2020,R2021}. The specific time duration for each events was selected to include the potential frequencies where we expect foreshock ULF waves \citep[see,][]{H1982,Sh2020}. More specifically, we looked for foreshock waves based on the magnetic field magnitude and the expected range of frequencies for these waves at Mars' heliocentric distance \citep[see Figure 1 in,][]{H1982}.}

To investigate the presence of ultra-low frequency waves when MAVEN was magnetically connect to the Martian bow shock, we used a criterion based on those employed  in \citet{Ro2016} and \citet{A2020}. For each event,
\begin{enumerate}
    \item We computed the total magnetic power spectral density (PSD) as a function of the frequency. 
    \item We defined an event with ULF wave activity if the PSD showed an increase within a frequency interval centered around a specific frequency $f_0$ compared to two adjacent windows each with a width {\color{black}$\Delta f = 0.2 f_0$}. Given the potential variability in the wave frequency ($f_0$) \citep[see,][]{R2020}, we applied the following condition for each frequency between 0.007 and 0.090 Hz,
    \begin{equation} \displaystyle \label{eq:condition}
        {\sum}\text{PSD}\left[{\bf B}(f)\right]|^{1.2f_{0}}_{0.8f_{0}} > {\sum}\text{PSD}[{\bf B}(f)]|^{1.4f_{0}}_{1.2f_{0}}, {\sum}\text{PSD}[{\bf B}(f)]|^{0.8f_{0}}_{0.6f_{0}}.
    \end{equation}
    \item Finally, we identified the frequency $f_0$ where the above condition was maximized, i.e., where the ratio of integrals as a function of the frequency $f_0$ was maximum, {\color{black} and we {denote} this frequency as $f_{id}$.}
\end{enumerate}
{\color{black} As we discussed in the Introduction, it is important to note that the identified wave frequency $f_{id}$ may falls within the PCWs frequency range in the Martian foreshock. We shall back to this point in Section \ref{subsec:freq}. Also, it is worth mentioning that we used a width of contiguous windows $\Delta f = 0.2 f_0$, a typical value for waves identification \citep[][]{Ro2016,A2020,Romeo2021}. However, when we range $\Delta f$ between $0.1 f_0$ and $0.3 f$, we observed a negligible impact on our statistical results.}

To estimate the polarization and propagation direction of these foreshock ULF waves, we employed the minimum variance analysis (MVA) \citep{S1967}. Generally, the MVA is used to estimate the normal direction to an approximately one-dimensional current layer, wave front, or other transition layer in a plasma. The detailed derivation of this method can be found elsewhere \citep[e.g.,][]{S1998}. Here, we summarize the main steps to find the MVA basis and its consequences on the propagation and polarization of a given planar wave. The MVA basis can be constructed using the auto-correlation magnetic matrix,
\begin{align}
    M_{pq}^B \equiv \langle B_p B_q \rangle - \langle B_p \rangle \langle B_q \rangle,
\end{align}
where $p$ and $q$ are the three magnetic field components. The auto-correlation matrix $M_{pq}^B$ is then diagonalized to obtain the three eigenvalues $\lambda_i$ ($i=1,2,3$) and the corresponding eigenvectors $\hat{\bf e}_i$. The eigenvector $\hat{\bf e}_3$ (associated with the smallest eigenvalue $\lambda_3$) serves as an estimator for the normal propagation vector $\hat{\bf k}$ to the planar wave front, with $\lambda_3$ representing the variance of the magnetic field component along the estimated normal. The eigenvectors $\hat{\bf e}_1$ and $\hat{\bf e}_2$, corresponding to the maximum and intermediate eigenvalues, respectively, indicate the tangential directions to the wave front. Additionally, the quasi-planar assumption of the wave train can be characterized by the $\lambda_2/\lambda_3$ ratio, which should be larger for planar waves. The polarization can be characterized by the $\lambda_1/\lambda_2$ ratio. For example, ratios near 1 are indicative of nearly circularly polarized waves \citep{S1998}.

\begin{figure}
\begin{center}
\includegraphics[scale=.95]{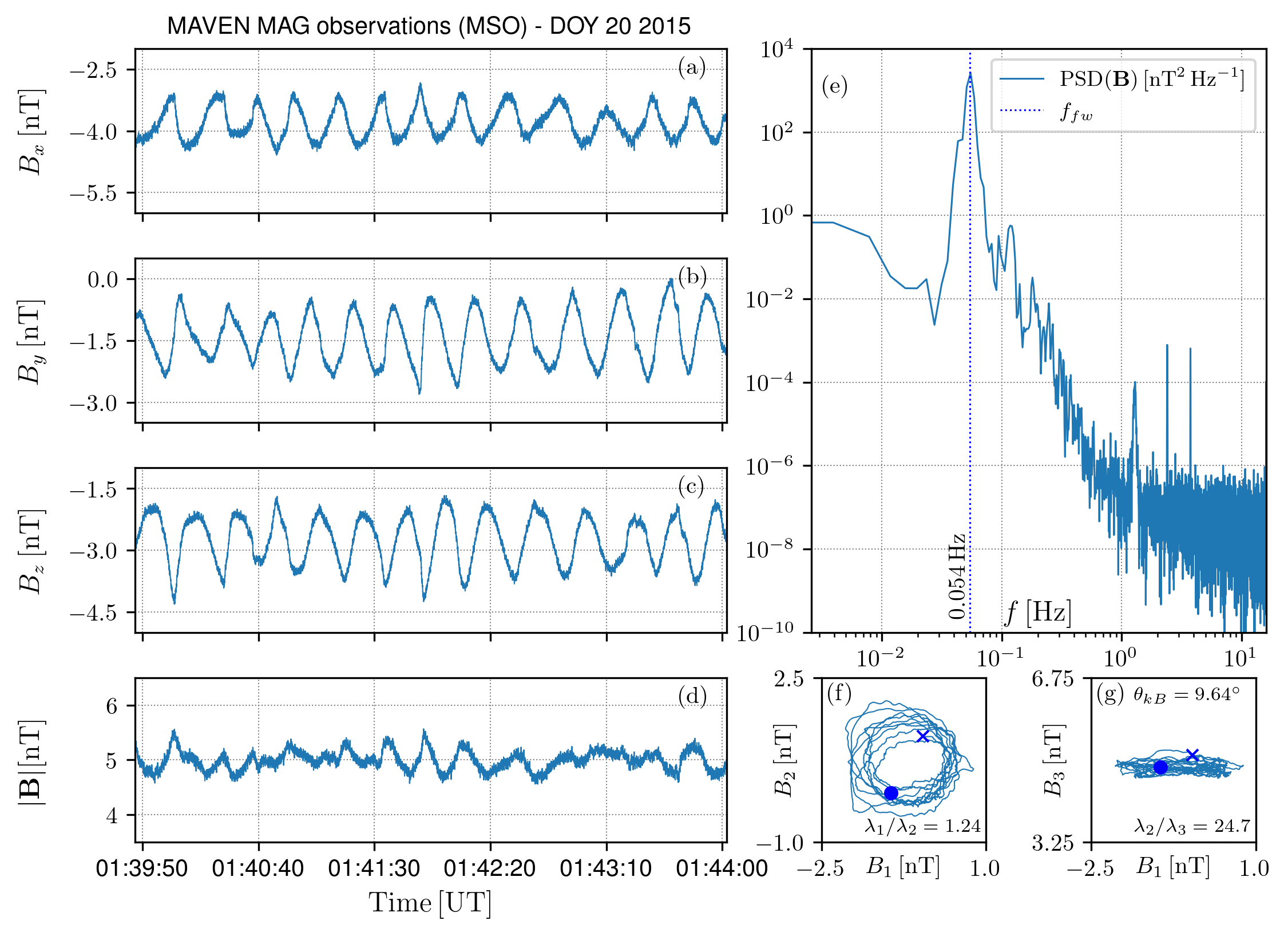}
\end{center}
\caption{ULF foreshock wave event seen by MAVEN on DOY 20 2015 at 01:39:46.85 UT. (a-d) Magnetic field components in MSO and absolute value as a function of time. (e) Magnetic PSD as a function of the frequency $f$. Hodograms for (f) the intermediate component $B_2$ as a function of the maximum component $B_1$ and (g) the minimum component $B_3$ as a function of $B_1$. It is important noting that the mean $B_3$ component is negative.}
\label{fig1}
\end{figure}

\section{Foreshock ULF waves: a case study on January 20, 2015} \label{sec:case}

Figure \ref{fig1} presents a foreshock ULF wave event identified using the detection criteria described in Section \ref{sec:obs}. Figure \ref{fig1} (a-d) show the magnetic field components and magnitude as a function of time for a 255-s event seen by MAVEN on DOY 20 2015 at 01:39:46.85 UT. The magnetic field components are shown in the Mars-centered Solar Orbital (MSO) coordinates, where the $\hat{\bf x}$ direction points toward the Sun, the $\hat{\bf z}$ direction is perpendicular to Mars’s orbital plane (positive toward the ecliptic north) and the $\hat{\bf y}$ axis completes the right-handed system. The three magnetic ﬁeld components exhibit oscillations with a well-deﬁned frequency. Figure \ref{fig1} (e) shows the PSD for the total magnetic field as a function of the frequency. Our method identified a clear peak at 0.0541 Hz, corresponds to the foreshock wave frequency $f_{fw}$. It is noteworthy that the PCW frequency $f_{ci}$ for this event is 0.0758 Hz, determined by the magnetic field magnitude (approximately $\sim$ 5 nT) since $f_{ci} = q|{\bf B}|/m$. Additionally, a slight increase in power at $f\sim1$ Hz, indicating the presence of whistler waves upstream of the planetary bow shock, commonly referred to as “1 Hz” waves \citep[see, e.g.,][]{J2024}. These waves are typically found in planetary foreshocks, suggesting the employed selection criteria are appropriate.

Figure \ref{fig1} (f-g) display the hodograms from the MVA analysis for a specific sub-interval, where (f) shows the intermediate component $B_2$ as a function of the maximum component $B_1$, and (g) shows the minimum component $B_3$ as a function of $B_1$. It is important to note that the mean $B_3$ component is positive. Considering the clock-wise sense of gyration (blue cross towards blue dot) of the magnetic field fluctuations relative to ${\bf B}_0 = (-0.53,0.62,4.82)$ nT, the MVA analysis suggests that the polarization is left-handed, in the spacecraft reference frame. Moreover, these foreshock waves exhibit circular polarization (indicated by the small $\lambda_1/\lambda_2$) and a well-defined basis (indicated by the large $\lambda_2/\lambda_3$). The results are consistent with a quasi-parallel propagation with respect to the background magnetic field ${\bf B}_0$ ($\theta_{kB}=9.64^\circ$). 

The main properties presented here are in agreement with the characteristics of the so-called 30-s waves observed at Earth's foreshock \citep[][and references therein]{E2005,A2015}. These findings also concur with previous observational results on planetary foreshocks \citep[see, e.g.,][and references therein]{H1982,E2005,A2013,Sh2020,R2020,R2021}.

\begin{figure}
\begin{center}
\includegraphics[width=0.6\textwidth]{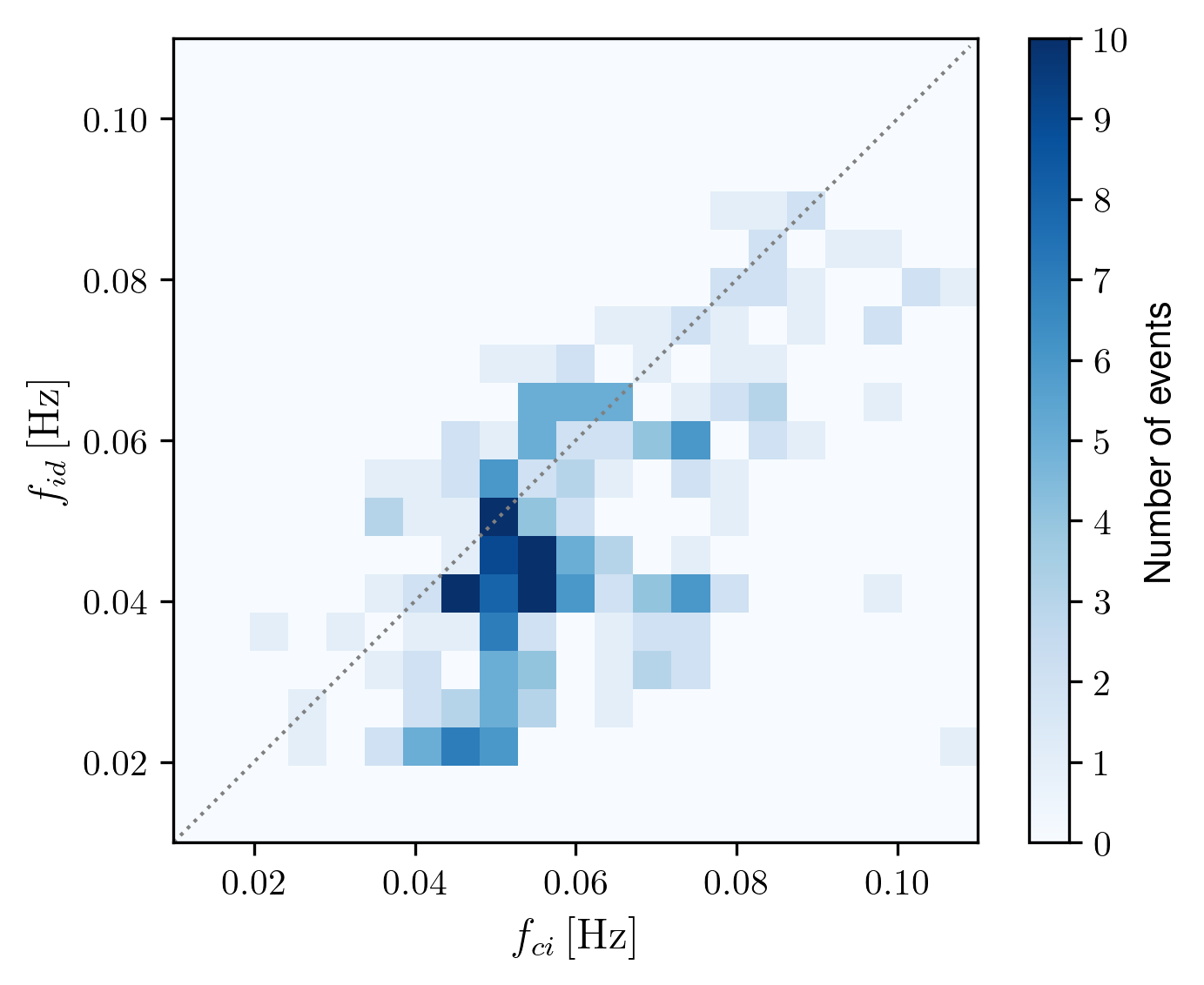}
\end{center}
\caption{{\color{black} Identified frequencies $f_{id}$ as a function of PCW frequencies $f_{ci}$. The colorbar corresponds to the number of events per bin.}}
\label{fig2}
\end{figure}

\begin{figure}
\begin{center}
\includegraphics[width=0.6\textwidth]{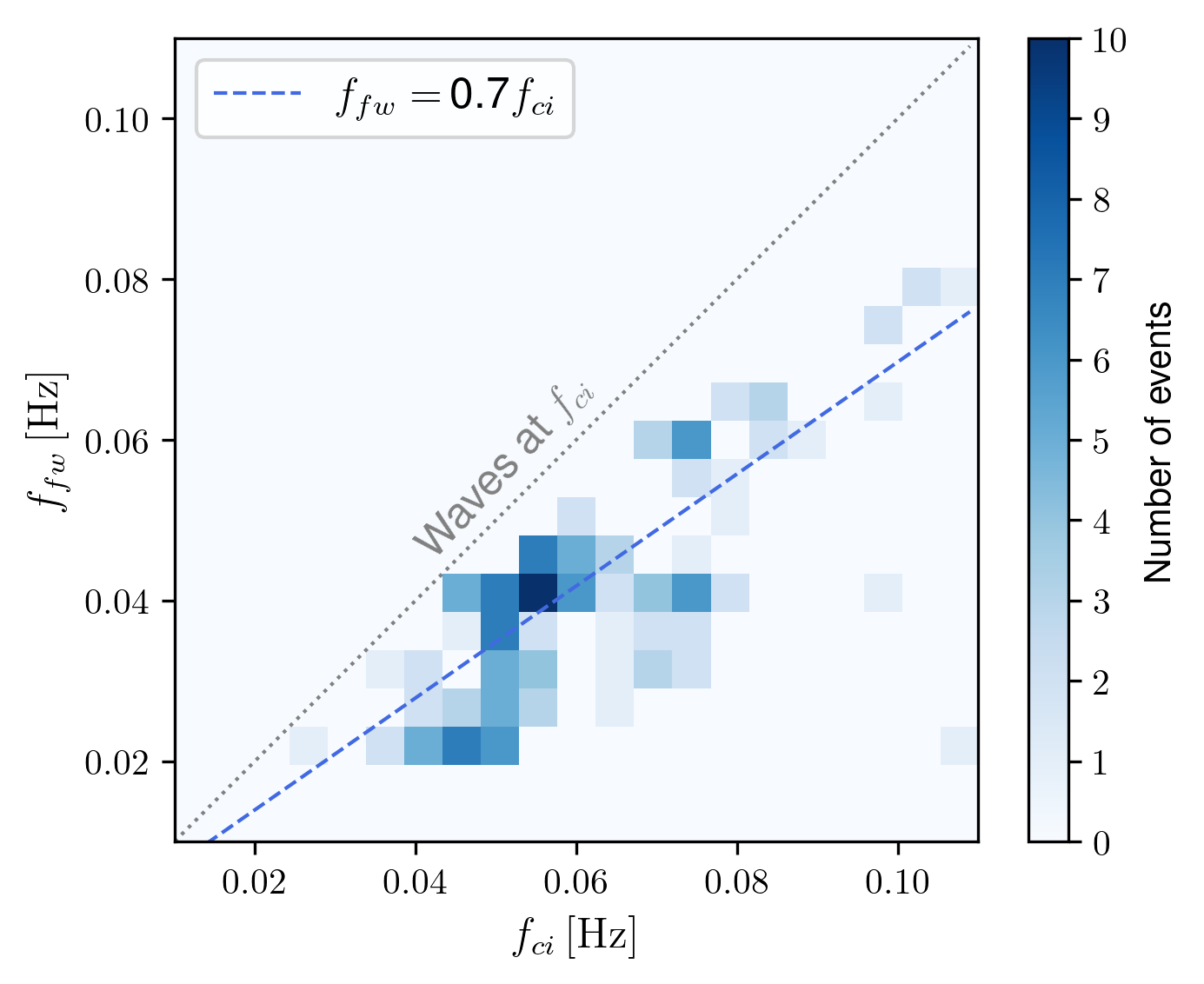}
\end{center}
\caption{{\color{black} Foreshock ULF wave frequencies $f_{fw}$ as a function of PCW frequencies $f_{ci}$. The dot-gray line corresponds to $f_{fw}=f_{ci}$. The dashed-blue line corresponds to the best linear fit for the foreshock ULF wave events.}}
\label{fig3}
\end{figure}


\section{Statistical Results} \label{sec:stat}

\subsection{\color{black} Foreshock ULF wave frequencies} \label{subsec:freq}

{\color{black} Figure \ref{fig2} shows the frequencies identified $f_{id}$ as a function of PCW frequencies $f_{ci}$, for all events. The dotted gray line represents $f_{id}=f_{ci}$. As we discussed in the Introduction, in contrast with other planets, such as Mercury \citep[e.g.,][]{R2021} or Earth \citep[e.g.,][]{A2015}, the Martian foreshock exhibits the coexistence of at least two types of waves. One type includes the ULF foreshock waves analyzed in the present work, while the other consists of proton cyclotron waves, which are observed at frequencies close to the local proton cyclotron frequency $f_{ci}$. Since these PCWs present wave properties similar to those of the ULF foreshock waves \citep[see,][]{H1982,R2021}, it is challenging to identifying  them using single spacecraft techniques. Specifically, within the region bounded by $f_{id} \leq 0.85f_{ci}$ \citep[see][]{D2008}, it is unclear whether we observe PCWs in the foreshock region or ULF foreshock waves. Therefore, {to ensure that we are studying only foreshock ULF waves, we excluded those events from our dataset.}

Figure \ref{fig3} shows distribution of the frequencies identified as foreshock ULF wave frequencies $f_{fw}$ as a function of PCW frequencies $f_{ci}$, after excluding potential PCW activity, that is, events with frequencies larger than 0.85$f_{ci}$. The best linear fit is $f_{fw}=\alpha f_{ci}$, with $\alpha=0.70\pm0.12$.}


\begin{figure}
\begin{center}
\includegraphics[width=0.85\textwidth]{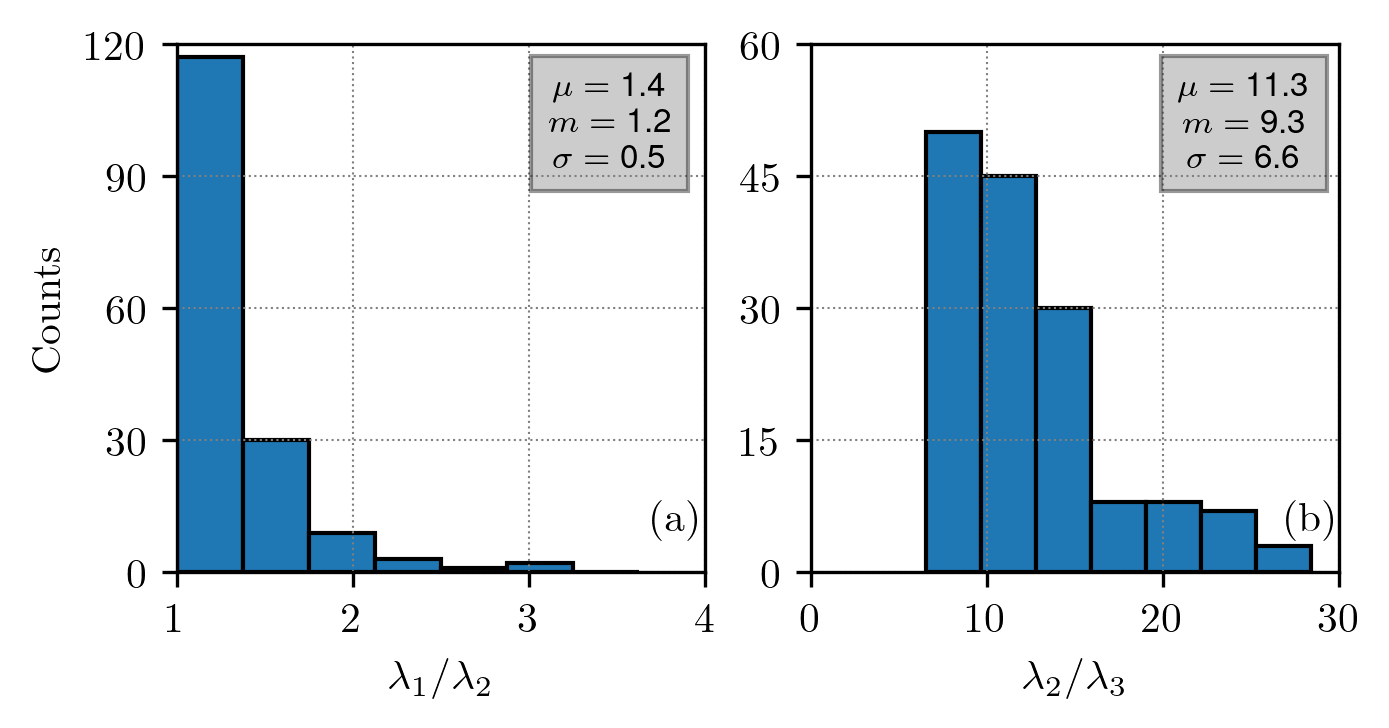}
\includegraphics[width=0.85\textwidth]{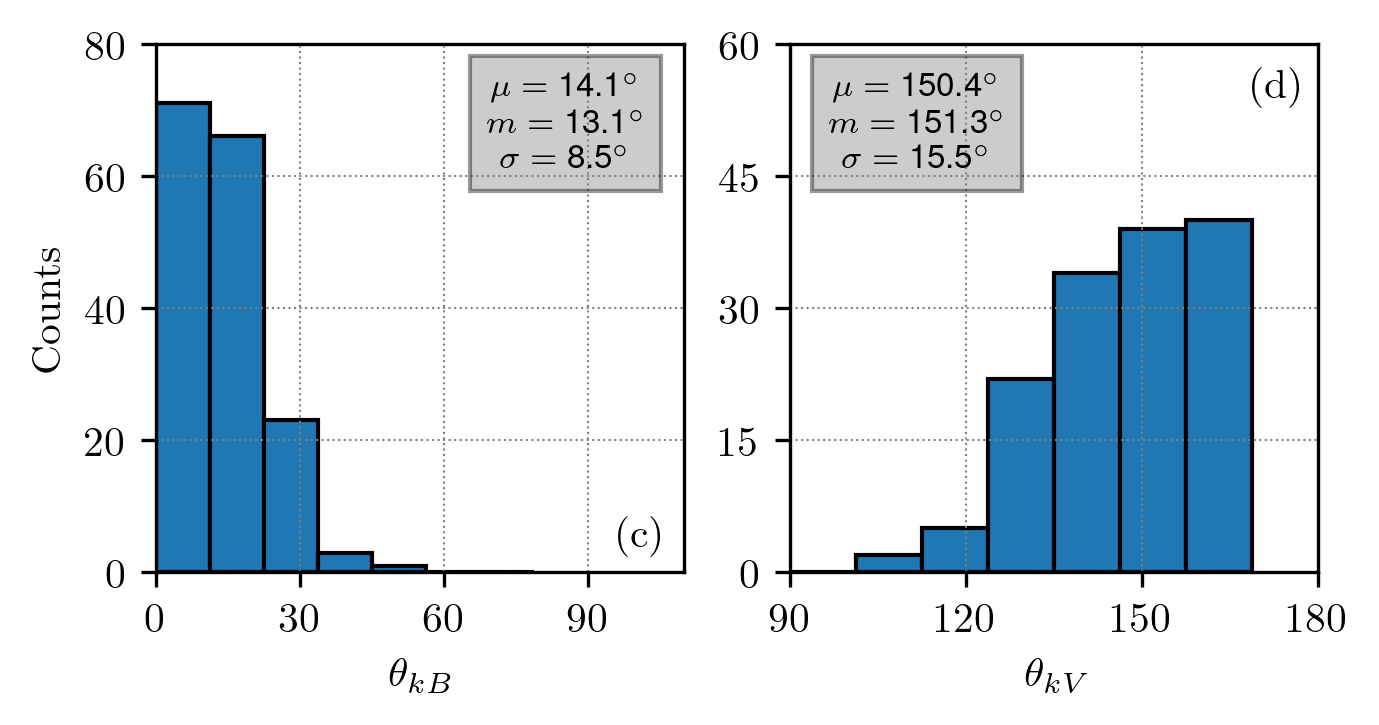}
\end{center}
\caption{Distributions of the main results from the minimum variance analysis for all the identified wave events: (a) maximum to intermediate and (b) intermediate to maximum eigenvalue ratios. Propagation angles of the wave front with respect to (c) the background magnetic field and (d) the mean velocity field, respectively.}
\label{fig4}
\end{figure}

\subsection{Minimum Variance Analysis Results} \label{subsec:mva}

Using the MVA described in Section \ref{sec:obs}, we conducted a statistical analysis of the properties of ULF waves in the Martian foreshock. Figure \ref{fig4} (a-b) show the maximum to intermediate (i.e., $\lambda_1/\lambda_2$) and the intermediate to minimum (i.e., $\lambda_2/\lambda_3$) ratios, respectively, for the identified ULF wave events. To ensure the presence of quasi-planar waves, we retained only those events with $\lambda_2/\lambda_3>5$  \citep[e.g.,][]{S1998,R2020}. Additionally, the distribution of $\lambda_1/\lambda_2$ ratios ranging between 1 and approximately 3 suggests the presence of nearly circularly {and elliptically} polarized waves. Furthermore, throughout most of the analyzed time intervals, all the studied cases exhibited left-handed polarization in the spacecraft reference frame.


Assuming that the identified cases are indeed quasi-planar and quasi-circularly/elliptically polarized waves, we can estimate the propagation angles with respect to the ambient magnetic field and the mean velocity field. These angles can be estimated using the following expressions,
\begin{align}
    \sin\theta_{kB} &= \frac{B_3}{|{\bf B}_0|}, \\
    \sin\theta_{kV} &= \hat{\bf k}\cdot\hat{\bf x},
\end{align}
where it is assumed that the solar wind inflow velocity is mainly in the $\bf\hat x$ MSO component. Figure \ref{fig4} (c-d) show the distribution of the angles $\theta_{kB}$ and $\theta_{kV}$, respectively. The mean ($\mu$), median ($m$) and standard deviation ($\sigma$) of these angles are listed in the first two rows of Table \ref{tab1}. As it was discussed in \citet{R2020}, we have assumed that direction $\bf\hat k$ points upstream.

\setlength{\tabcolsep}{6pt}
\renewcommand{\arraystretch}{1.1}
\begin{table}
    \centering
    \begin{tabular}{|c|ccc|}
    \hline
         & Mean ($\mu$)  & Median ($m$) & Std. ($\sigma$) \\
    \hline
       $\theta_{kB}$ [$^\circ$] & 13.9 & 13.1 & 7.9 \\
       $\theta_{kV}$ [$^\circ$] & 152.3 & 155.4 & 15.2 \\ 
    \hline
       $V_r/V_{sw}$  & 1.33  & 1.25 & 0.4 \\
       $V_{gc}/V_{sw}$  & 2.2  & 2.16 & 0.43 \\
    \hline
    \end{tabular}
    \caption{Mean ($\mu$), median ($m$) and standard deviation ($\sigma$) of the angles $\theta_{kB}$ and $\theta_{kV}$, and the velocity ratios $V_r/V_{sw}$ and $V_{gc}/V_{sw}$, respectively.}
    \label{tab1}
\end{table}

\begin{figure}
\begin{center}
\includegraphics[width=0.85\textwidth]{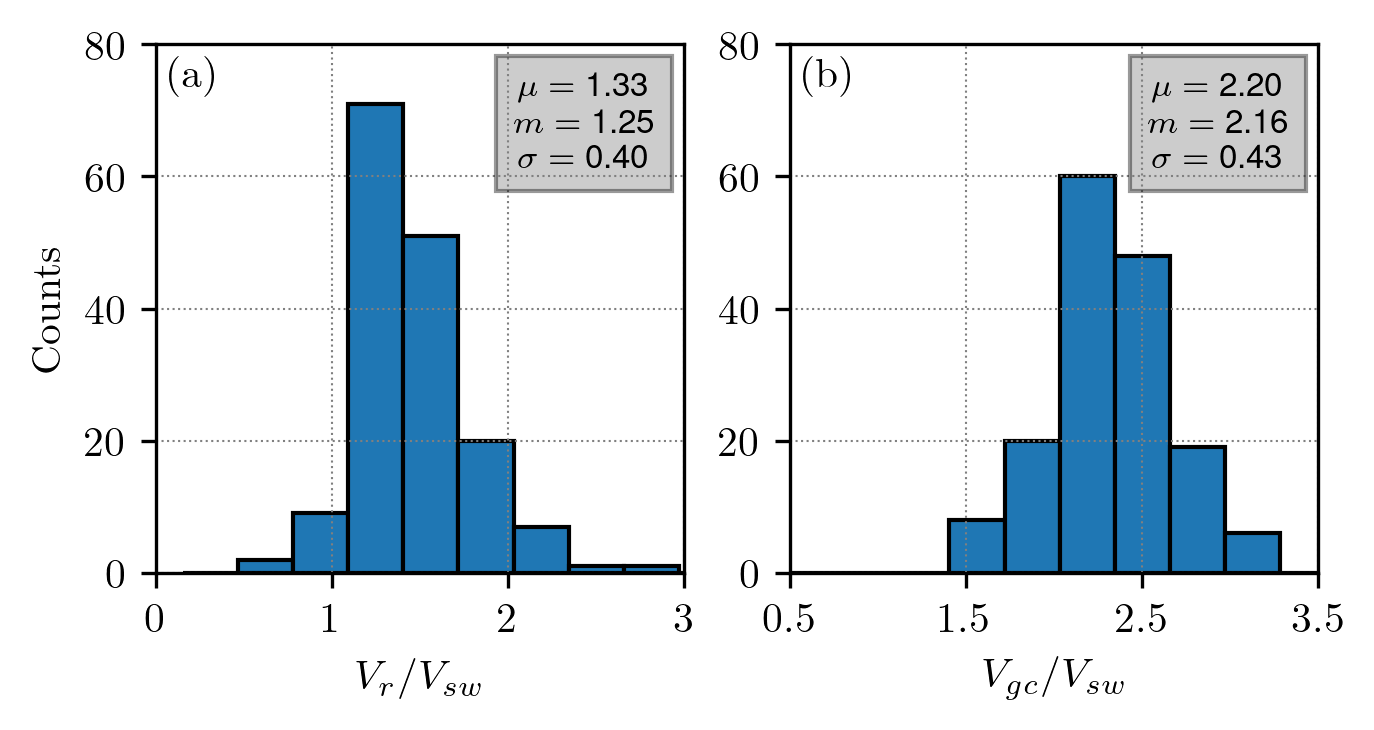}
\end{center}
\caption{Distribution of (a) $V_r/V_{sw}$ and (b) $V_{gc}/V_{sw}$ for all the ULF wave identified events.}
\label{fig5}
\end{figure}

\subsection{Particle acceleration mechanisms} \label{subsec:accel}

The foreshock ULF wave properties discussed in the previous Sections can be used to estimate properties of the backstreaming ions responsible for generating these waves \citep{G1993,M1998,M2003,M2007}. Specifically, two commonly used parameters can be estimated, the ratio of the particle velocity parallel to the mean magnetic field to the solar wind velocity ($V_r/V_{sw}$) and the ratio of the guiding center velocity of a backstreaming particle to the solar wind velocity ($V_{gc}/V_{sw}$). For a detailed derivation of these parameters, the reader is referred to \citet{M1998} and \citet{R2020} and the references therein. Briefly, using the cyclotron resonance condition between a backstreaming particle and a linear right-hand wave (in the plasma reference frame) and the Doppler shift equation between the plasma and the spacecraft reference frames, both parameters can be expressed as,
\begin{align}
    \frac{V_r}{V_{sw}} &= \left(\frac{1+a}{f_{fw}/f_{ci}-a}\right)\frac{\cos\theta_{kV}}{\cos\theta_{kB}} \\
    \frac{V_{gc}}{V_{sw}} &= \left[1+\left(\frac{V_r}{V_{sw}}\right)^2-2\left(\frac{V_r}{V_{sw}}\right)\cos\theta_{Bx}\right]^{1/2}
\end{align}
where $\theta_{Bx}$ is the angle between the ambient magnetic field and the mean solar wind velocity, {\color{black} and $a \equiv \omega / \Omega_{p}$, where $\omega$ is the wave frequency in the solar wind rest frame and $\Omega_p$ is the proton gyro-frequency. While the ${V_r}/{V_{sw}}$ ratio is derived in the plasma reference frame, the ${V_r}/{V_{sw}}$ ratio is expressed in the spacecraft reference frame.}

{\color{black} Figure \ref{fig5} (a-b) shows the distributions of $V_{r}/V_{sw}$ and $V_{gc}/V_{sw}$, respectively. Our statistical analysis shows that the mean values are $V_r/V_{sw} = 1.33\pm0.40$ and $V_{gc}/V_{sw}=2.20\pm0.43$.  It is worth mentioning that we used a value close to what is observed in the terrestrial foreshock $a=0.15$. However, it should be noted that the mean, median, and standard deviation of $V_r/V_{sw}$ and $V_{gc}/V_{sw}$ do not show strong variability when we range $a$ between 0.05 and 0.25 \citep[see also,][]{R2020}. Our observational results fall within the range of previously reported ratios for different planetary foreshocks \citep[see,][]{H1982,M1998,A2013,A2015,Sh2018,R2020,R2021}.}

\section{Discussion and Conclusions} \label{sec:diss}

To the best of our knowledge, our study presents the first statistical analysis of ultra-low frequency waves at the Martian foreshock. Using over two years of data from the Magnetometer instrument onboard the MAVEN spacecraft, we focused on periods when MAVEN was magnetically connected to Mars' bow shock. By applying a specific criterion for wave identification, we determined that the foreshock ULF waves exhibit moderate wave amplitudes and frequencies typically ranging from approximately 0.008 Hz to 0.086 Hz. Based on the MVA, these waves predominantly display circular or elliptical left-handed polarization in the spacecraft reference frame, consistent with a distribution of $\lambda_1/\lambda_2$ ratios between 1 and approximately 3. 

Using magnetic field and plasma observations from MAVEN, \citet{Sh2020} identified a foreshock wave event with a wave period of around 25 s, corresponding to roughly twice the local proton cyclotron period. They reported a large wave amplitude, a propagation angle $\theta_{kB} \sim 34^{\circ}$ and a left-handed elliptical polarization with respect to the ambient magnetic field (in the spacecraft frame). Our statistical study of 288 identified ULF waves events supports and generalizes these previous observations, enabling a comprehensive investigation of ULF wave properties. It is noteworthy that the observed frequency spectrum falls within frequencies derived from global hybrid models \citep[see,][]{J2022}. Specifically, the authors identified frequencies around 0.008 Hz, corresponding to a far region of the foreshock and frequencies around 0.08 Hz, related to a near foreshock region,  consistent with ULF waves traveling downstream through the bow shock. Additionally, they found left-handed polarization in the simulation frame (planetary frame) in the near and far region. Our observed frequencies predominantly fall in between the far and the near regions of the foreshock, with left-handed polarization noted in all studied events. However, a complete investigation of the spatial distribution of ULF waves is beyond the scope of this work and it is part of a future study taking into account the foreshock coordinates \citep[see,][]{Greenstadt1986,A2013,A2015}. 


As discussed in the Introduction, the Martian foreshock contains ULF waves and another type of wave known as proton cyclotron waves, which share similar frequency range and polarization properties. Unfortunately, with the techniques and observations used in this study, we cannot completely exclude the presence of PCW activity in our analysis. However, it is important to note that excluding events with frequencies within the expected range for PCWs (i.e., foreshock frequencies between $f_{fw} = 0.85f_{ci}$ and $f_{fw}=1.15f_{ci}$), does not significantly affect the statistical properties of the waves or their implications for the acceleration models. {\color{black} After removing potential PCWs events, the slope $\alpha$ between the foreshock ULF wave frequency $f_{fw}$ and the PCW frequency $f_{ci}$ is $(0.7\pm0.15)$.} \citet{R2020} conducted a statistical analysis of ULF wave activity in Mercury's foreshock using high-resolution magnetic field measurements from MESSENGER. The authors found properties similar to those analyzed here and identified a clear relationship between the observed frequency $f_{fw}$ and the local magnetic field magnitude $B$. Specifically, they determined that $f_{fw}(\text{Hz})=(0.00796\pm 0.00170)~|B|~(\text{nT})$, which corresponds to $f_{fw}=(0.52 \pm 0.11)f_{ci}$. Notably, their findings agree within uncertainties with earlier observational results reported by \citet{H1982}. 
 

Using the foreshock ULF wave properties obtained from the minimum variance analysis, we estimated the characteristics of the backstreaming ions responsible for generating the ULF waves \citep[e.g., see][]{G1993}. In particular, we estimated the ratio between the particle velocity parallel to the mean magnetic field and the solar wind velocity $V_{r}/V_{sw}$, and the ratio between the guiding center velocity of a backstreaming particle and the solar wind velocity $V_{gc}/V_{sw}$. In the past, these two parameters have been used to limit the potential mechanism present in planetary bow shocks \citep[see,][]{H1982,M1998,A2013,A2015,Sh2018,R2020,R2021}. For the Earth's foreshock, \citet{M1998} and \citet{A2015} reported $\langle V_{gc}/V_{sw}\rangle=1.05\pm 0.01 (\theta_{Bx}=45^\circ)$,  $\langle V_{gc}/V_{sw}\rangle=1.11\pm 0.04$ and $\langle V_{gc}/V_{sw}\rangle=1.68$ (for the field-aligned beam boundary), respectively. More recently, \citet{R2020} reported $\langle V_{gc}/V_{sw}\rangle=0.76\pm 0.27$ for Mercury, and \citet{Sh2018} reported $\langle V_{gc}/V_{sw}\rangle=1.07~(\theta_{Bx}=36^\circ)$ for Venus. Our results are compatible with these previous observational findings. However, we are obtaining slightly larger values of $\langle V_{gc}/V_{sw}\rangle$ due to the probable inclusion of mix contributions from different cone angle. For the $V_{r}/V_{sw}$ ratio, previously observational results are reported as 1.2-2.2 for Mercury, 1.7/1.9 for Venus, 2.5$\pm$0.3 for Earth and 2.1/2.3 for Jupiter \citep[see,][]{H1982,R2020}. Our distribution of $V_{r}/V_{sw}$ encompasses these previous observational results and is consistent with resonant conditions for the backstreaming distributions with velocities varying between approximately 0.5 and 2 times the solar wind velocity. The similarity of this range to previous planetary foreshock observations may indicates the presence of an universal acceleration process acting in planetary shocks and responsible for this particular energetic backstreaming population. Finally, it is worth mentioning that a deep and conclusive analysis on the possible acceleration mechanisms responsible for the backstreaming distributions necessarily involves the use of particle observations in the Martian foreshock and is beyond the scope of the present work. 


Our results reveal that Martian ULF foreshock waves share similarities with those observed at other planets, showing a potential universality on their characteristics. The observed linear relationship between the frequency and the magnetic field strength further corroborates the idea that these waves are produced by resonant instabilities, despite the presence of newborn planetary ions and PCWs. Our assessment of backstreaming proton energy, aligning with values found in other planetary foreshocks, indicates that gyro-resonance instabilities significantly contribute to wave generation. This raises the intriguing question of whether one type of plasma instability might dominate over another within the turbulent plasma environment \citep{Ru2017,Halekas2020_waves,A2020,A2021,Romanelli2022APJ,R2024,R2024b}. Mars, with its unique heliospheric position and extended hydrogen exosphere, provides an exceptional platform for investigating these phenomena, offering insights into additional plasma instabilities that could overshadow traditional planetary foreshock instabilities. These findings emphasize the significance of Mars as a natural laboratory for advancing our understanding of planetary foreshock dynamics and their broader implications for space plasma physics.

\begin{acknowledgments}
The MAVEN project is supported by NASA through the Mars Exploration Program.  {\color{black} N.A.~acknowledges financial support from the following grants: PIP Grant No.~11220200101752, UBACyT Grant No.~ 20020220300122BA and Redes de Alto Impacto REMATE from Argentina.} N.R.~is supported through a cooperative agreement with Center for Research and Exploration in Space Sciences $\&$ Technology II (CRESST II) between NASA Goddard Space Flight Center and University of Maryland, College Park, under award number 80GSFC24M0006. {\color{black}The MAVEN data are publicly available through the Planetary Plasma Interactions Node of the Planetary Data System (https://pds-ppi.igpp.
ucla.edu/). MAVEN MAG data \citep{MAG_data} was used.}

\end{acknowledgments}

\bibliographystyle{aasjournal}

\end{document}